\documentclass[supp]{tlp}
\usepackage{aopmath}
\usepackage{supp}

\makeatletter

\def\define#1{\@ifnextchar [{\@MYargdef#1}{\@MYargdef#1[0]}}

\def\@MYargdef#1[#2]#3{\@ifdefinable #1{\@MYreargdef#1[#2]{#3}}}

\def\redefine#1{\edef\@tempa{\expandafter\@cdr\string 
  #1\@nil}\@ifundefined{\@tempa}{\@latexerr{\string#1\space undefined}\@ehc
    }{}\@ifnextchar [{\@MYreargdef#1}{\@MYreargdef#1[0]}}

\catcode`\?=11\relax
\def\@MYreargdef#1[#2]#3{\@tempcnta#2\relax\let#1\relax
\edef\@tempa{\def#1}\@tempcntb \@ne
\let\@?@?\relax\@whilenum\@tempcnta>0
\do{\edef\@tempa{\@tempa\@?@?\the\@tempcntb}\advance\@tempcntb \@ne \advance
\@tempcnta \m@ne}\let\@?@?##\@tempa{#3}}
\catcode`\?=12\relax

\makeatother

\renewcommand{\bmod}%
{\mskip-\medmuskip \mkern5mu \mathbin{\idFont mod} \mkern5mu \mskip-\medmuskip}

\define{\mathsym}[1]{\relax\ifmmode#1\else
	\errmessage{Mathematical symbol outside math mode}\fi}
\define{\idFont}{\sf}
\define{\mathid}[1]{\mathsym{{\idFont #1}}}

\define{\m}[1]{$#1$}
\define{\M}[1]{$$#1$$}

\define{\gt}{\mathsym{\mathchar 12606\relax}} 
\define{\lt}{\mathsym{\mathchar 12604\relax}} 

\define{\paren}[1]{(#1)}
\define{\parenA}[1]{(#1)}
\define{\parenB}[1]{\bigl(#1\bigr)}
\define{\parenC}[1]{\Bigl(#1\Bigr)}
\define{\parenAuto}[1]{\left(#1\right)}
\define{\setOf}[1]{\lbrace#1\rbrace}
\define{\setA}[1]{\lbrace#1\rbrace}
\define{\setB}[1]{\bigl\lbrace#1\bigr\rbrace}
\define{\setC}[1]{\Bigl\lbrace#1\Bigr\rbrace}
\define{\setCond}[2]{\lbrace#1\mathrel{\vert}#2\rbrace}
\define{\setCondA}[2]{\lbrace#1\mathrel{\vert}#2\rbrace}
\define{\setCondB}[2]{\bigl\lbrace#1\bigm\vert#2\bigr\rbrace}
\define{\setCondC}[2]{\Bigl\lbrace#1\Bigm\vert#2\Bigr\rbrace}

\define{\unionUntil}{\union\cdots\union}
\define{\unionMulti}[2]{\bigcup_{#1}^{#2}}
\define{\intersect}{\cap}
\define{\intersectUntil}{\intersect\cdots\intersect}
\define{\intersectMulti}[2]{\bigcap_{#1}^{#2}}
\define{\timesUntil}{\times\cdots\times}
\define{\setsize}[1]{{\left\vert #1 \right\vert}}
\define{\compl}[1]{\overline{#1}}

\define{\powerset}[1]{2^{#1}}
\define{\inB}{\;\in\;}

\define{\defEq}{:=}
\define{\defEqB}{\;:=\;}
\define{\defIff}{\;\mathsym{:\Longleftrightarrow}\;}
\define{\metaThen}{\mathsym{\;\Longrightarrow\;}}
\define{\metaIf}{\mathsym{\;\Longleftarrow\;}}
\define{\metaIff}{\mathsym{\;\Longleftrightarrow\;}}

\define{\lneg}{\mathsym{\neg}}
\define{\lif}{\mathsym{\leftarrow}}
\define{\lifDup}{\mathsym{\Leftarrow}}
\define{\lthen}{\mathsym{\rightarrow}}
\define{\liff}{\mathsym{\leftrightarrow}}
\define{\lfalse}{\mathid{false}}
\define{\ltrue}{\mathid{true}}

\define{\lorUntil}{\lor\cdots\lor}
\define{\landUntil}{\land\cdots\land}
\define{\lorMulti}[2]{\bigvee_{#1}^{#2}}
\define{\landMulti}[2]{\bigwedge_{#1}^{#2}}

\define{\lfor}[4]{#1\:#2\:#3:#4}
\define{\lexistsQ}{\exists}
\define{\lexists}[3]{\lfor{\lexistsQ}{#2}{#1}{#3}}
\define{\lallQ}{\forall}
\define{\lall}[3]{\lfor{\lallQ}{#2}{#1}{#3}}

\define{\answer}{\mathid{answer}}

\define{\natNum}{\mathsym{{\rm l\kern-0.13em N}}}

\define{\realNum}{\mathsym{{\rm I\kern-0.14em R}}}

\define{\struct}[1]{\langle#1\rangle}
\define{\twoCases}[4]{\left\{\begin{array}{l@{\kern10pt}l}
				#1&\mbox{#2}\\#3&\mbox{#4}
				\end{array}\right.}
\define{\until}{, \ldots,}
\define{\seqOf}[1]{#1^*}
\define{\emptySeq}{\epsilon}
\define{\w}{\mathid{w}}

\define{\code}[1]{{\tt #1}} 

{\begin{center}\begin{tt}\begin{tabular}{@{}l@{}}}%
{\end{tabular}\end{tt}\end{center}}

	{\begin{center}\begin{tt}\begin{tabular}{@{}l@{ }l@{}}}%
	{\end{tabular}\end{tt}\end{center}}

\define{\U}{{\char95}} 
\define{\LT}{{\char60}} 
\define{\GT}{{\char62}} 
\define{\B}{{\char92}} 
\define{\AMP}{{\char38}} 
\define{\D}{{\char36}} 
\define{\SN}{{\char126}} 
\define{\HASH}{{\char35}} 
\define{\Q}{{\char34}} 
\define{\PCT}{{\char37}} 
\define{\LB}{{\char123}} 
\define{\RB}{{\char125}} 
\define{\HAT}{{\char94}} 

\define{\ALPH}{\mathid{ALP\kern-0.08em H}}
\define{\LOG}{\mathid{LOG}}
\define{\VARS}{\mathid{V\kern-0.17em ARS}}
\define{\var}{\mathsym{X}}
\define{\varA}{\mathid{X}}
\define{\varB}{\mathid{Y}}
\define{\varC}{\mathid{Z}}
\define{\const}{\mathsym{c}}
\define{\constA}{\mathsym{a}}
\define{\constB}{\mathsym{b}}
\define{\constC}{\mathsym{c}}
\define{\data}{\mathsym{d}}
\define{\dataSet}{\mathsym{{\cal D}}}
\define{\sig}{\Sigma}
\define{\SORTS}{\mathid{{\cal S}}}
\define{\sort}{\mathid{s}}
\define{\PREDS}{\mathid{{\cal P}}}
\define{\pred}{\mathid{p}}
\define{\predA}{\mathid{p}}
\define{\predB}{\mathid{q}}
\define{\predC}{\mathid{r}}
\define{\FUNS}{\mathid{{\cal F}}}
\define{\fun}{\mathid{f}}
\define{\args}{\alpha}
\define{\argsB}{\beta}
\define{\argSorts}{\alpha}
\define{\resSort}{\rho}
\define{\interp}{\mathid{{\cal I}}}
\define{\interpB}{\mathid{{\cal J}}}
\define{\ass}{\mathid{{\cal A}}}
\define{\iV}[2]{(#1,#2)}
\define{\eval}[2]{#1\lbrack\kern-0.15em\lbrack#2\rbrack\kern-0.15em\rbrack}
\define{\evalV}[3]{\eval{\iV{#1}{#2}}{#3}}
\define{\varDecl}{\nu}
\define{\TERMS}{\mathid{T\kern-0.1em E}}
\define{\term}{\mathid{t}}
\define{\termB}{\mathid{u}}
\define{\AT}{\mathid{AT}}
\define{\FO}{\mathid{FO}}
\define{\fo}{\varphi}
\define{\foB}{\psi}
\define{\fos}{\Phi}
\define{\modify}[3]{#1\langle#2/#3\rangle}
\define{\impl}{\vdash}
\define{\apply}[2]{#2{}#1}
\define{\subst}{\theta}
\define{\substB}{\sigma}
\define{\mgu}{\mathid{mgu}}
\define{\doSubst}[2]{#2\,#1}
\define{\HU}{\mathid{{\cal U}}}
\define{\HB}{\mathid{{\cal B}}}
\define{\HSet}{\mathid{H}}
\define{\ground}{\mathid{ground}}
\define{\LAT}{\mathid{{\cal M}}}
\define{\LATB}{\mathid{{\cal N}}}

\define{\emptyClause}{{\hbox{%
	\setlength{\unitlength}{0.24ex}%
	\begin{picture}(5,5)(0,0)
	\put(0,0){\line(0,1){5}}
	\put(0,0){\line(1,0){5}}
	\put(0,5){\line(1,0){5}}
	\put(5,0){\line(0,1){5}}
	\end{picture}}}}

\define{\lit}{\mathid{L}}
\define{\litA}{\mathid{A}}
\define{\litB}{\mathid{B}}
\define{\litC}{\mathid{C}}
\define{\Body}{\mathsym{{\cal B}}}
\define{\Rule}{\mathsym{{\cal R}}}
\define{\F}{\mathid{F}} 
\define{\ruleFun}[1]{\mathid{r}_{#1}}

\define{\prog}{\mathsym{{\idFont P}}}
\define{\DB}{\mathsym{{\idFont D}}}
\define{\Base}{\mathsym{{\cal B}}}
\define{\SEM}{\mathsym{{\idFont S}}}
\define{\T}[1]{\mathsym{{\idFont T}}_{#1}}
\define{\TP}{\T{\prog}}
\define{\norm}{\mathsym{{\idFont norm}}}
\define{\Tneg}[2]{\mathsym{{\idFont T}}_{#1,#2}}
\define{\lub}{\mathid{lub}}
\define{\glb}{\mathid{glb}}
\define{\lfp}{\mathid{l\kern-0.1em f\kern-0.1em p}}

\define{\I}{\mathid{{\cal I}}}
\define{\J}{\mathid{{\cal J}}}

\define{\nf}{\mathop{\mathid{not\kern0.2em}}}
\define{\nfPred}[1]{\mathop{\mathid{not{\U}}#1}}

\define{\free}{\mathid{f}}
\define{\bound}{\mathid{b}}

\define{\bp}{\mathid{bp}}
\define{\binding}{\beta}
\define{\bindingSet}{{\cal B}}

\define{\vars}{\mathsym{{\cal X}}}
\define{\varsA}{\mathsym{{\cal X}}}
\define{\varsB}{\mathsym{{\cal Y}}}
\define{\varsC}{\mathsym{{\cal Z}}}

\define{\inputVars}{\mathid{input}}

\define{\freeVar}{\mathid{vars}}
\define{\varsOf}{\mathid{vars}}
\define{\boundPos}{\mathid{bound}^+}
\define{\boundNeg}{\mathid{bound}^-}
\define{\unboundPos}{\mathid{unbound}^+}
\define{\unboundNeg}{\mathid{unbound}^-}
\define{\err}{\mathid{err}}

\define{\gbpPos}{\mathid{gbp}^+}
\define{\gbpNeg}{\mathid{gbp}^-}
\define{\intersectGBP}{\mathid{intersect}}
\define{\unionGBP}{\mathid{union}}

\define{\transPos}{\mathid{trans}^+}
\define{\transNeg}{\mathid{trans}^-}
\define{\transS}{\mathid{trans}^\mathid{s}}
\define{\idPos}{\mathid{id}^+}
\define{\idNeg}{\mathid{id}^-}
\define{\idNil}{\mathid{nil}}
\define{\idLeft}{\mathid{left}}
\define{\idRight}{\mathid{right}}
\define{\idCons}{\mathid{cons}}
\define{\idVar}{\mathid{ID}}
\define{\idFun}{\mathid{id}}
\define{\idRule}{\mathid{rule}}
\define{\idPred}{\mathid{pred}}

\define{\path}{\mathid{path}}
\define{\edge}{\mathid{edge}}

\define{\query}{\mathsym{Q}}
\define{\state}{\mathsym{S}}
\define{\states}{\mathsym{{\cal S}}}
\define{\goals}{\mathsym{{\cal G}}}
\define{\goal}{\mathsym{g}}
\define{\goalSet}{\mathsym{G}}
\define{\answers}{\mathsym{{\cal A}}}
\define{\dedRelE}{\mathsym{\mapsto_\epsilon}}
\define{\dedRelD}[1]{\mathsym{\mapsto_{#1}}}
\define{\SYS}{\mathsym{{\cal T}}}

\define{\varV}{\mathid{V}}

\define{\sg}{\mathid{sg}}
\define{\person}{\mathid{person}}
\define{\X}{\mathid{X}}
\define{\Xp}{\mathid{Xp}}
\define{\Y}{\mathid{Y}}
\define{\Yp}{\mathid{Yp}}

\define{\emptyC}{{\setlength{\unitlength}{0.6mm}%
		\begin{picture}(3,3)(0,0)
			\put(0,0){\line(1,0){3}}
			\put(0,0){\line(0,1){3}}
			\put(0,3){\line(1,0){3}}
			\put(3,0){\line(0,1){3}}
		\end{picture}}}

\define{\cl}[1]{\mathid{cl}_{#1}}

\define{\grandparent}{\mathid{grandparent}}
\define{\parent}{\mathid{parent}}
\define{\father}{\mathid{father}}
\define{\mother}{\mathid{mother}}
\define{\personA}{\mathid{ann}}
\define{\personB}{\mathid{betty}}
\define{\personC}{\mathid{chris}}
\define{\personD}{\mathid{doris}}
\define{\inputPred}{\mathid{input}}

\define{\C}{\mathid{C}}

\begin{document}

\title[A Framework for Bottom-Up Simulation of SLD-Resolution]
	{A Framework for Bottom-Up Simulation of SLD-Resolution}
\author[S.~Brass]{STEFAN BRASS\\
	Martin-Luther-Universit\"at Halle-Wittenberg,
	Institut f\"ur Informatik,\\
	Von-Seckendorff-Platz~1, D-06099 Halle (Saale), Germany\\
	\email{brass@informatik.uni-halle.de}}

\pagerange{\pageref{firstpage}--\pageref{lastpage}}
\volume{\textbf{10} (3)}
\jdate{March 2002}
\setcounter{page}{1}
\pubyear{2002}

\pubauthor{Brass}
\jurl{xxxxxx}
\pubdate{22 June 2013}

\submitted{14 February 2014}
\revised{---}
\accepted{---}

\maketitle

\label{firstpage}


\begin{abstract}
This paper introduces a framework for the bottom-up
simulation of SLD-resolution based on partial evaluation.
The main idea is to use database facts to represent a
set of SLD goals. For deductive databases it is natural
to assume that the rules defining derived predicates are known
at ``compile time'', whereas the database predicates are known
only later at runtime.
The framework is inspired by the author's own SLDMagic method,
and a variant of Earley deduction
recently introduced by Heike Stephan and the author.
However,
it opens a much broader perspective.
\end{abstract}


\section{Introduction}
\label{secIntro}

Deductive databases 
use logic programming
for data intensive applications.
For example,
database queries are written in a Prolog-like language
called Datalog.
Basic Datalog is pure Prolog without structured terms.
The data stored e.g.~in a relational database
can be seen as a large set of facts.
Of course,
many extensions of this basic language have been investigated
and implemented in prototype systems.
While in the beginning,
the main achievement of deductive databases
was seen in the possibility to write recursive queries,
e.g.~for hierarchical and graph-structured data,
the more general goal
is to support database queries and application programming
in one declarative language.

For efficient query evaluation,
it is important to distinguish
between predicates defined by rules in a logic program,
and predicates defined by facts in the database.
The logic program is known at ``compile time'',
while the facts are known only at ``runtime''.
I.e.~the database facts form the input to the logic program.
Because the program might be executed several times
with different database states,
it pays off to invest time for optimization
by precomputing as much as possible given the logic program,
while the database facts are not yet known.
Time might be saved even in a single execution of the program,
because the logic program is usually 
small,
while the database state is big.

Deductive databases use bottom-up evaluation,
i.e.~apply the 
\m{T_P}-operator
to derive facts from already known facts
to get logically implied instances of the query.
Of course,
many optimizations are added to this basic method.
Especially,
there are a lot of methods for making bottom-up evaluation
goal-oriented,
i.e.~to derive only facts that are in some sense needed
for computing answers to the query.
Most well-known in this area is the ``magic set'' method~\cite{BMSU86},
where magic sets simply encode subqueries.

Fran{\c c}ois Bry had the idea to explain magic sets
with a meta-interpreter,
which describes top-down evaluation,
but runs on a bottom-up machine~\cite{Bry90}.
When this meta-interpreter is partially evaluated
with respect to the given rules,
one gets exactly the result of the magic set transformation.
This might be the first case of partial evaluation
for strict bottom-up evaluation,
other applications of partial evaluation for deductive databases are,
e.g.~\cite{SI88,LMK90,Han95}.
Of course,
partial evaluation for top-down evaluation (Prolog)
was well investigated
(see, e.g.,~\cite{Gal93}),
but the methods for partial evaluation depend crucially
on the execution model.

It turned out that magic sets do not exactly correspond
to SLD-resolution,
and that in the case of tail recursions,
SLD-resolution has a big advantage,
because it does not need to materialize every proven ``lemma''
\cite{Ros91,Bra95magic}.

The author then proposed a meta-interpreter
that describes SLD-resolution exactly
and runs on a bottom-up machine.
In this way,
set-oriented evaluation techniques can be used,
and termination can be guaranteed for Datalog,
e.g.~a rule like
\m{\predA(\varA)\lif\predA(\varA)}
does not cause an infinite loop.
In contrast to magic sets and tabling techniques,
tail-recursion runs much faster,
e.g.~consider the following logic program~\m{\prog}:
\begin{quote}
\m{\begin{array}{@{}lcl@{}}
\path(\varA,\varB)&\:\lif\:&\edge(\varA,\varB).\\
\path(\varA,\varC)&\:\lif\:&\edge(\varA,\varB)\land\path(\varB,\varC).
\end{array}}
\end{quote}
Suppose the database is \m{\DB\defEq\setCond{\edge(i-1,i)}{1\le i\le n}},
i.e.~the graph is a single path of length~\m{n}.
For the query~\m{\path(0,\varA)},
the SLD-tree has \m{4n+3}~nodes,
i.e.~a number that is linear in~\m{n},
whereas magic sets derive all facts of the form \m{\path(i,j)}
with \m{0\le i\lt j\le n},
which is quadratic in~\m{n}.
While a tail-recursion optimization for magic sets
has already been studied in~\cite{Ros91},
our ``SLDMagic''-method~\cite{Bra00} had other advantages as well
by simulating SLD-resolution bottom-up.
It passes
also non-equality conditions on parameters
to called predicates,
and avoids joins when a predicate ``returns''.


However,
the possibilities for bottom-up simulation of SLD-resolution
extend much farther
if we allow a single fact to represent multiple nodes in the SLD-tree.
Already when the SLDMagic method was implemented,
it was noted
that it produces a lot of ``copy'' rules,
which only copy tuples from one predicate to another predicate.
These rules were then eliminated by a postprocessing step,
which merged the predicates.
This can be understood as allowing facts to represent
a set of goals in the SLD tree.

Recently,
Heike Stephan and the author developed a variant of Earley deduction
\cite{PW83,Por86},
also tabling \cite{TS86,CW96SLG,NC12} can be seen as developing
the Earley method further.
Our Earley deduction variant
uses states describing a relatively big part of deduction
using only program rules, but no database facts~\cite{BS13}.
While there are differences in the technical details,
this can be seen as representing a whole set of SLD goals
in a single ``state'',
which is encoded a a database fact.

The purpose of this paper is to introduce an abstract framework for simulating
SLD-resolution on a bottom-up machine,
improve the understanding of the mentioned methods and their similarities,
and discuss options for improving the efficiency
of program execution in deductive databases.

\section{Basic Definitions}
\label{secBasicDefinitions}

A logic program~\m{\prog} is a finite set of rules of the form
\m{\litA\lif\litB_1\landUntil\litB_n}
where \m{\litA} and \m{\litB_i}, \m{i=1\until n} (with \m{n\ge0})
are positive literals, i.e.~have the form \m{\pred(\term_1\until\term_m)}
with a predicate~\m{\pred} and terms~\m{\term_j}, \m{j=1\until m}.
Terms are variables or constants.
We assume that the rules are range-restricted (safe),
i.e.~every variable appearing in the head~\m{\litA}
also appears in the body~\m{\litB_1\landUntil\litB_n}.
In this paper, we do not consider negation.

A subset of the predicates are selected as EDB predicates
(``extensional database'').
These are the database relations.
The EDB predicates can appear in the logic program only in the body
(i.e.~in the~\m{\litB_i}),
but not in the head (\m{\litA}).
Besides the logic program,
a database state~\m{\DB} is given,
which is a finite set of facts (positive ground literals)
in which only EDB~predicates appear
(it follows that \m{\prog} and~\m{\DB} are disjoint).
We write \m{\Base} for the set of all positive ground literals
with EDB predicate
(the Herbrand base restricted to EDB predicates).
This set will usually be infinite
(because it permits arbitrary strings, integers, etc.~as arguments).
Of course,
\m{\DB\subseteq\Base}.


The predicates which do appear in rule heads
are called IDB predicates (``intensional database'').
These predicates are defined by means of rules,
not by enumerating facts.
It is possible that an IDB predicate has only rules with empty body
(i.e.~facts),
then the only difference to an EDB predicate
is that we assume that the program is given for partial evaluation
(``compile time''),
whereas the database is only known at runtime.

In practice,
one also needs ``built-in predicates'' like~\m{\lt},
which are defined by procedures inside the system.
Such predicates raise interesting questions of range restriction and safety,
see, e.g.,~\cite{RBS87,KRS88,Bra09RangeRestriction}.
However,
to simplify the presentation,
we exclude them here.

A query (goal)~\m{\query} is a conjunction~\m{\litA_1\landUntil\litA_n}
of positive literals (like a rule body).
The variables appearing in the query are called the answer variables.
The purpose of query evaluation is to find ground substitutions
for the answer variables
such that the corresponding instance of the query is true in
the minimal model of~\m{\prog\union\DB}.

To simplify the presentation,
we assume the first literal selection rule for SLD-resolution
(as applied in Prolog).
In SLD-resolution,
the computed answer substitution is normally defined
by looking at an entire SLD-derivation leading to the empty clause:
\begin{quote}
\m{\query\longrightarrow\goal_1\longrightarrow\goal_2\longrightarrow\;\cdots\;
	\longrightarrow\goal_n\longrightarrow\emptyC}.
\end{quote}
But it suffices to consider only single goals at a time,
if we start with an ``extended query'',
which has a literal with all answer variables
and a special predicate~\m{\answer} at the very end:
\kern0.2em
\m{\litB_1\landUntil\litB_n\land\answer(\varA_1\until\varA_m)}.
The predicate~\m{\answer} is not otherwise used
in the program or the database.
It only does the bookkeeping of the current values
of the answer variables
(since all substitutions done during the SLD derivation
are also applied to this literal).
Therefore,
if we reach a goal consisting of
a single literal~\m{\answer(\const_1\until\const_m)},
we know that the answer substitution
\m{\setOf{\var_1/\const_1\until\var_m/\const_m}}
has been computed.

It might seem at first that it is a restriction
that we assume a 
completely given query.
One might want to consider also ``parameterized queries'',
containing constants not yet known at ``compile time''.
For instance,
one might want to do query optimization (partial evaluation)
for any query of the form~\m{\path(\const,\varA)},
with an arbitrary constant~\m{\const}.
This corresponds to ``binding pattern''~\m{\mathid{bf}} (mode~\m{\mathid{+-}})
as used in the magic set method.
However,
we can simply use the query
\m{\mathid{input}(\mathid{C})\land\path(\mathid{C},\varA)}
with a database predicate~\m{\mathid{input}}
which is filled with the parameter value before query evaluation starts.




\section{States Representing Sets of Goals}
\label{secStates}

\subsection{Definition of SLDDB-Systems}

We first define an abstract system with states representing sets of
goals in the SLD tree,
and a transition relation between these states.
The states will later be encoded as facts,
and the transition relation corresponds to rules
which permit to derive these facts.

A SLDDB-System~\m{\SYS} consists of
\begin{itemize}
\item
\m{\states}, a (usually infinite) set of states,
\item
\m{\goals(\state)}, a non-empty set of goals for every \m{\state\in\states},
\item
\m{\state_0\in\states}, an initial state,
\item
\m{\dedRelE\;\subseteq\states\times\states},
a relation between states (\m{\epsilon}-transitions),
\item
\m{\dedRelD{\F}\;\subseteq\states\times\states},
a relation between states for every possible database fact~\m{\F\in\Base}.
\end{itemize}

Given an SLDDB-System~\m{\SYS} as above,
and a database state~\m{\DB\subseteq\Base} (set of facts),
an answer tuple \m{(\const_1\until\const_m)} is computed
iff there is a finite sequence~\m{\state_0\ldots\state_n\;\in\states^*}
of states,
starting with the initial state~\m{\state_0},
containing the answer in the final state,
i.e.~\m{\answer(\const_1\until\const_m)\in\goals(\state_n)},
and such that for~\m{i=1\until n},
\begin{itemize}
\item
\m{\state_{i-1}\dedRelE\state_i}
or
\item
\m{\state_{i-1}\dedRelD{\F}\state_i} for some~\m{\F\in\DB}.
\end{itemize}

The states will later be encoded as facts,
and the reachable states will be computed bottom-up,
therefore repeated states in a state sequence can be detected
in order to improve termination.
Of course, if an analysis shows that this cannot happen,
one can save the effort for duplicate detection.
Another useful property is that for certain states,
there is only one possible successor state
(if \m{\DB} satisfies integrity constraints such as keys).

\subsection{Correctness of SLD-Simulation}

An SLDDB-System should correspond to SLD-resolution
for a given program and query.
First we define the correctness,
i.e.~that all goals occurring in a state sequence
are really derivable from the query, the program,
and the database facts used in state transitions.

Let a logic program~\m{\prog} and a query~\m{\query} be given.
An SLDDB-System~\m{\SYS} is correct with respect to~\m{\prog} and~\m{\query}
iff
\begin{itemize}
\item
For every~\m{\goal\in\goals(\state_0)},
there is an SLD-derivation of~\m{\goal} from the extended query
\m{\query\land\answer(\varA_1\until\varA_m)}
using rules in~\m{\prog}
(this includes the case that the derivation is empty,
i.e.~\m{\goal} is the extended query).
\item
Whenever \m{\state_1\dedRelE\state_2},
then for every \m{\goal_2\in\goals(\state_2)}
there is \m{\goal_1\in\goals(\state_1)}
such that there is a non-empty
SLD-derivation~\m{\goal_1\longrightarrow\goal'\longrightarrow^*\goal_2}
of~\m{\goal_2} from~\m{\goal_1},
using rules in~\m{\prog},
and the first step~\m{\goal'} in this derivation
is contained in~\m{\goals(\state_2)}, i.e.~\m{\goal'\in\goals(\state_2)}.
\item
Whenever \m{\state_1\dedRelD{\F}\state_2},
then for every \m{\goal_2\in\goals(\state_2)}
there is \m{\goal_1\in\goals(\state_1)}
such that there is a non-empty
SLD-derivation~\m{\goal_1\longrightarrow\goal'\longrightarrow^*\goal_2}
of~\m{\goal_2} from~\m{\goal_1},
where the first step uses the fact~\m{\F},
other steps use rules in~\m{\prog},
and \m{\goal'\in\goals(\state_2)}.
\end{itemize}

If this condition is satisfied,
then for every computed answer tuple~\m{(\const_1\until\const_m)}
there is an SLD-derivation of~\m{\answer(\const_1\until\const_m)}
from \m{\query\land\answer(\varA_1\until\varA_m)}.
This obviously means that there is also an SLD-derivation
of the empty clause from~\m{\query},
where the
substitution~\m{\subst\defEq\setOf{\varA_1/\const_1\until\varA_m/\const_m}}
is applied to the variables in~\m{\query}.
Therefore,
by the correctness of SLD-resolution,
it follows that \m{\doSubst{\subst}{\query}}
is a logical consequence of~\m{\prog\union\DB}.

\subsection{Completeness of SLD-Simulation}

In the opposite direction,
we need that every SLD-derivation is indeed represented
in the states and the transition relation.
Note that the SLDDB-System is independent of a concrete database state,
i.e.~it represents derivations using any possible database facts.
Of course,
when a state sequence is constructed to answer a query in a concrete
database state~\m{\DB},
only facts in~\m{\DB} can be used.

Let a logic program~\m{\prog} and a query~\m{\query} be given.
An SLDDB-System~\m{\SYS} is complete with respect to~\m{\prog} and~\m{\query}
iff
\begin{itemize}
\item
The extended query~\m{\query\land\answer(\varA_1\until\varA_m)}
is contained in~\m{\goals(\state_0)}.
\item
For every state~\m{\state\in\states} and every goal~\m{\goal\in\goals(\state)}
and every~\m{\goal'} which can be derived from~\m{\goal}
and a rule in~\m{\prog} by an SLD-resolution step,
there is a variant~\m{\goal''} of~\m{\goal'}
with \m{\goal''\in\goals(\state)}
or \m{\goal''\in\goals(\state')} for some state~\m{\state'}
with \m{\state\dedRelE\state'}.
\item
For every state~\m{\state\in\states}, every goal~\m{\goal\in\goals(\state)}
and every~\m{\goal'} which can be derived from~\m{\goal}
and a fact~\m{\F\in\Base}
by an SLD-resolution step,
there is a variant~\m{\goal''} of~\m{\goal'}
and state~\m{\state'} with \m{\state\dedRelD{\F}\state'}
such that \m{\goal''\in\goals(\state')}.
\end{itemize}

This condition ensures that every SLD-derivation
of the empty clause from~\m{\query} using rules in~\m{\prog}
and facts in~\m{DB} 
can indeed be represented by a state sequence.

It would actually suffice to require the completeness
not for all SLD resolution steps,
but only for steps in successful derivations
where the set of facts used in the state sequence
satisfies given integrity constraints.
In this way, ``dead ends'' could be cut off early.

\subsection{Example: SLD-Resolution}

Of course,
one would expect that SLD-resolution itself fits in this framework.

If one wants to simulate SLD-resolution exactly,
including the non-termination for \m{\pred(\varA)\lif\pred(\varA)},
one uses SLD-derivations as states.
I.e., given a program~\m{\prog} and a query~\m{\query},
the set of states~\m{\states} is the set of SLD-derivations
\m{\query\land\answer(\varA_1\until\varA_m)\longrightarrow^*\goal_n}
and the set of goals for the above state~\m{\state}
is \m{\goals(\state)\defEq\setOf{\goal_n}},
i.e.~a singleton set consisting of the last (or current) goal
in the derivation.
The transition relations extend the derivation by one goal,
i.e.~lead from~\m{\state}
to the following state~\m{\state'}:
\begin{quote}
\m{\query\land\answer(\varA_1\until\varA_m)
	\longrightarrow^*\goal_n\longrightarrow\goal_{n+1}}.
\end{quote}
If in the last SLD resolution step
a program rule was used,
\m{\state\dedRelE\state'}.
If instead a fact~\m{\F} with EDB-predicate was used,
\m{\state\dedRelD{\F}\state'}.

\subsection{Example: SLD-Resolution Without Duplicate Nodes}

Of course,
it is more in the spirit of bottom-up evaluation
to eliminate duplicate nodes,
and this is what SLDMagic~\cite{Bra00} did.
In the above framework,
this simply means that the states are single goals,
i.e.~conjunctions of the form
\begin{quote}
\m{\litB_1\landUntil\litB_n\land\answer(\term_1\until\term_m)}.
\end{quote}
Furthermore,
we have to exclude goals which differ only in a variable renaming.
We do this by requiring that variables are named~\m{\varV_1,\varV_2,\ldots}
in the order of first occurrence in the goal.
Let \m{\norm(\goal)} be a mapping from goals to goals
which normalizes variables in this way.
Of course,
when states are single goals,
we can simply let~\m{\goals(\state)\defEq\setOf{\state}}.
The transition relation is simply SLD-resolution plus the normalization.
I.e.~\m{\state\dedRelE\state'} holds iff
\m{\state'} is derivable from~\m{\state}
by a single SLD resolution step (using a rule in~\m{\prog}),
followed by variable normalization.
Correspondingly, \m{\state\dedRelD{\F}\state'} holds
when fact~\m{\F\in\Base} is used in the resolution step.
Obviously, the set of computed answers is not changed
by merging nodes in the SLD-tree with the same goal.
The number of duplicate answers is changed
(every distinct answer is computed only once).
However,
if one considers duplicates as important,
one probably wants a more declarative specification.
Duplicates in deductive databases have been considered e.g.~in~\cite{MPR90}.

Note also that this works only because in SLD-resolution
there is no need to return to the ``caller''
--- otherwise the same subgoal could appear in different contexts,
and it might be important to distinguish between them.
This was one of the difficulties in our variant of Earley deduction.
In SLD-resolution, the entire continuation of the proof is built into the goal.
With the \m{\answer}-literal at the end of the goal,
even for determining the answer substitution,
we do not have to look at an entire path in the tree.

Termination can be guaranteed if duplicate states in state sequences
are excluded and the program is at most tail-recursive,
i.e.~only the last literal of a rule can be a recursive call.
This property ensures that the length of goals is bounded~\cite{Bra00}.

\subsection{Maximal States}

\label{subsecMaxStates}

So far,
states contained only single goals.
But 
we want to compute as much as possible
at ``compile time'',
i.e.~when query and program are known,
but the facts in the database are not yet known.
Therefore,
we do the following closure operation
on sets of goals:
\begin{quote}
\m{\cl{\prog}(\goalSet)\defEq\setCond{\norm(\goal')}{
	\mbox{there is \m{\goal\kern-0.2pt\in\kern-0.2pt\goalSet}
		and an SLD-derivation
		\m{\goal\longrightarrow^*\goal'}
		using only rules in~\m{\prog}}}}.
\end{quote}
Now,
given program~\m{\prog} and query~\m{\query}
with answer variables~\m{\varA_1\until\varA_m},
the initial state is the closure of the extended version of~\m{\query}:
\begin{quote}
\m{\state_0\defEq
	\cl{\prog}\parenB{\setOf{\query\land\answer(\varA_1\until\varA_m)}}}.
\end{quote}
Given a state~\m{\state} and a fact~\m{\F} with EDB-predicate,
the successor state~\m{\state'} (with \m{\state\dedRelD{\F}\state'})
is defined as follows:
\begin{quote}
\m{\state'\defEq
	\cl{\prog}\parenB{\setCond{\goal'}{\mbox{there is \m{\goal\in\state}
			such that a single SLD-resolution step
			of~\m{\goal} and~\m{\F} gives~\m{\goal'}}}}}.
\end{quote}
The other transition relation~\m{\dedRelE} is empty,
i.e.~state transitions occur only when EDB-facts are used,
other deductions (with program rules)
are done within the states.

Of course,
it is possible that states become infinite.
For instance,
consider the left recursive version of the transitive closure example:
\begin{quote}
\m{\begin{array}{@{}lcl@{}}
\path(\varA,\varB)&\:\lif\:&\edge(\varA,\varB).\\
\path(\varA,\varC)&\:\lif\:&\path(\varA,\varB)\land\edge(\varB,\varC).
\end{array}}
\end{quote}
Let the query be~\m{\path(0,\varA)}.
Then the initial state contains
\begin{quote}
\m{\begin{array}{@{}l@{}}
\path(0,\varV_1)\land\answer(\varV_1).\\
\edge(0,\varV_1)\land\answer(\varV_1).\\
\path(0,\varV_1)\land\edge(\varV_1,\varV_2)\land\answer(\varV_2).\\
\edge(0,\varV_1)\land\edge(\varV_1,\varV_2)\land\answer(\varV_2).\\
\path(0,\varV_1)\land\edge(\varV_1,\varV_2)\land\edge(\varV_2,\varV_3)
	\land\answer(\varV_3).\\
\edge(0,\varV_1)\land\edge(\varV_1,\varV_2)\land\edge(\varV_2,\varV_3)
	\land\answer(\varV_3).\\
\ldots
\end{array}}
\end{quote}
However,
this is not necessarily a problem if one can work
with finite representations of this infinite set.
For instance,
the left recursive transitive closure works well
in our variant of Earley deduction~\cite{BS13},
and the graphs of partially processed rules
used there can be seen as encoding an infinite set of SLD goals
(if there are cycles in the ``called by'' relation).
%
Furthermore,
we have the following theorem,
which shows that for programs without left recursions,
the closure operation will not lead to infinite states:

\begin{theorem}
Suppose that \m{\prog} contains no IDB facts
(i.e.~all rules have a non-empty body),
and no left recursions,
i.e.~the predicate of the first body literal of each rule
does not depend on the predicate defined by the rule
(i.e.~it does not call --- possibly indirectly --- that predicate).
Then \m{\cl{\prog}(\goalSet)} is finite for every finite~\m{\goalSet}.
\end{theorem}

Suppose for the moment
that we can precompute all states
(with parameters for the constants from database facts
only known at runtime).
Then working with maximal states is in some sense
as efficent as it can get
(when we look only at a single, successful derivation):
One of the previously unknown database facts is processed in each step.
If none of them is redundant
(that depends on the program, e.g.~there should be no repeated subgoals),
any other query evaluation method must touch the same facts.
But methods like ``magic sets'' also generate a lot of rules
which do not contain EDB literals in the body.
These rules are applied in addition to the necessary deductions
with EDB literals.
Furthermore,
only a comparison with the magic set version with ``supplementary predicates''
would be fair
(otherwise there are repeated accesses to the same fact
in a derivation of a single answer).
But then even more intermediate literals derived.


So,
where is the hitch?
At the moment,
we can do the precomputation of states
only for certain programs.
Extending this set of programs is an interesting research problem.

Furthermore,
there is a fundamental difference between SLD resolution and magic sets,
namely,
SLD resolution proves IDB literals always in the context of a concrete call
(the goal contains everything that has to be done after the literal is proven),
whereas magic sets derive ID predicates in isolation.
Both has sometimes advantages:
SLD resolution saves joins to get the proven literal back into the context
of the caller,
and with a more interesting selection function,
conditions on the result can be checked at the best moment,
and sometimes this is before the call is fully finished,
see~\cite{Bra00}.
However,
when magic sets have proven an IDB literal,
they can use it several times in different contexts.
SLD resolution (and thus our approach) has to prove it repeatedly.
In the computation of a single answer this probably does not occur often,
but when all answers are needed as in deductive databases,
this might sometimes lead to suboptimal behaviour.
In~\cite{Bra00} we proposed to mix both approaches,
by doing explicit subproofs for certain literals.
The same technique would work here.
Ideally,
we would have an automatic decision which of the two methods is better
for a concrete call.
This remains a research problem, too.

\section{Encoding States as Facts}
\label{secEncoding}

Our goal now is to study possibilities
for encoding states as facts,
such that the transition relation can be computed with standard Datalog rules.
In this way,
the approach becomes a source-to-source transformation like magic sets.
However,
the resulting rules have a very simple structure,
such that other implementations,
like a direct translation to C{\tt ++},
are an interesting option.

Such an encoding is also required because there is normally
an infinite number of states:
The SLDDB-system models deductions with all possible database facts,
e.g.~containing arbitrary strings as arguments.
Often constants from the facts will be contained in the goals
for continuing the proof,
and therefore, the number of states is infinite.
But for the precomputation at compile time,
we need a finite representation of the set of states.

This is done by introducing parameterized states,
which are mappings from a certain number of data values
(the parameter values or arguments)
to states.
If we introduce a predicate for a parameterized state
(with the arity equal to the number of parameters),
we only have to define which concrete state
is represented by a fact with this predicate.

Another way to see this is that at ``compile time'',
we do not know the concrete constants from the database
(only constants occurring in the program).
Therefore,
we represent these unknown values by ``parameters''.
Later, at ``runtime'',
we have values for the parameters,
and can fill in the ``holes''
to get a fully specified state.

However,
simply replacing parameters in the goals
by constants is not the only way
how states can be encoded.
As explained below,
the counting method can be understood as encoding
the length of a part of a goal (of very regular structure)
in a parameter value.
This permits to handle (at least some) cases
where the length of the occurring goals depends on the data,
therefore, we cannot precompute them at compile time.
Note that this is different from the left recursive version
of the transitive closure discussed above:
There, we had goals of arbitrary length in a single state,
and therefore did not need to store any concrete length.
In the programs,
for which the counting method is made
(especially the same generation example),
the exact length is important.

\subsection{Parameters for Constants Known Only at Runtime}

\label{subsecParamsForConsts}

Let us first consider the case where the parameters are simply
replaced by constants.
We assume that there are special variables~\m{\C_1,\C_2,\ldots}
for the parameters
(disjoint from the variables~\m{\varV_1,\varV_2,\ldots}
used for normalization).
Now a parameterized state with \m{n}~parameters
is defined by a set of goals
in which the variables~\m{\C_1\until\C_n} can occur,
and the variables~\m{\varV_1,\varV_2,\ldots},
but no other variables.
Furthermore,
each goal must satisfy the normalization requirement,
i.e.~if the variable~\m{\varV_i}, \m{i\gt 1}, occurs somewhere in a goal,
all variables~\m{\varV_1\until\varV_{i-1}} must occur to the left
in the same goal.
Whereas the scope of standard variables~\m{\varV_1,\varV_2,\ldots}
is only a single goal,
i.e.~they are a kind of ``local variables'',
parameters are ``global'' in the parameterized state (set of goals).
Therefore,
a normalization is more difficult
(we must use a standard order of goals),
but of course,
we do not construct distinct states
which differ only in a renaming of the parameters.
Now, if a parameterized state~\m{\goalSet} with \m{n}~parameters
is encoded as predicate~\m{\pred},
then the fact \m{\pred(\const_1\until\const_n)}
represents
\m{\setCondB{\doSubst{\subst}{\goal}}{
	\goal\in\goalSet,\;
	\subst=\set{\C_1/\const_1\until\C_n/\const_n}}}.

The unification procedure must be changed in order to respect the parameters.
We must keep in mind that at runtime,
there will be constants 
for them.
The first change is that if we need to unify a parameter
and a standard variable,
we replace the variable by the parameter.
The second change is that when we need to unify two parameters,
or a parameter and a constant,
unification produces a condition for the parameter values,
e.g.~\m{\C_i=\constA}.
Thus, the unification procedure does not only yield a substitution
for the standard variables,
but also a (consistent) conjunction of conditions for the parameters.
For each call to the unification procedure,
we must make a case distinction.
Either the actual parameter values satisfy the condition
(e.g.~\m{\C_i=\constA}),
and the unification succeeds,
or they do not (\m{\C_i\ne\constA}), and the unification fails.
Note that when we work with sets of goals,
some failed unifications do not necessarily lead to the empty result set.
If the computation of the successor state needs \m{k}~unifications,
there could be \m{2^k}~cases to distinguish,
but usually it will be much less,
because we can stop as soon as the condition which describes the case
becomes inconsistent.
Each case might yield a different parameterized state.
In the rule that describes the state transition,
we have to check the conditions on the parameters,
and also the non-equality conditions
in order to avoid computing the same SLD derivation twice.
See also~\cite{BS13}.

\subsection{Other Encodings: Counting}

For general recursions,
goals might become larger and larger depending on the data,
thus it is not possible to precompute them explicitly at compile time
(even if we replace unknown constants by parameters).
E.g.,
this happens in the ``same generation'' example:
\begin{quote}
\m{\begin{array}{@{}lcl@{}}
\sg(\varA,\varA)&\lif&\mathid{person}(\varA).\\
\sg(\varA,\varB)&\lif&\parent(\varA,\varA')\land\sg(\varA',\varB')
			\land\parent(\varB,\varB').
\end{array}}
\end{quote}
However,
other types of encodings are possible.
For instance,
when applying the counting method~\cite{BMSU86,GZ92}
to the same generation example,
one can view \m{\mathid{c\_sg}(\C,\mathid{I})}
as representing
\begin{quote}
\m{\sg(\C,\varB_1)\land
\parent(\varB_2,\varB_1)\landUntil\parent(\varB_{\mathid{I}+1},\varB_{\mathid{I}})\land
						\answer(\varB_{\mathid{I}+1}).}
\end{quote}

\section{Conclusions}
\label{SecConclusions}

This paper offers a different view on some previous methods
for query evaluation,
such as SLDmagic, counting, and a variant of Earley deduction.
By introducing a common framework for them,
one can compare and combine their features,
and this also opens a space for thinking about new, improved methods.

Obviously,
there are currently still many (interesting) research questions,
and few readymade methods beyond what was already there.
Nevertheless,
the understanding ist improved,
and the potential of the presented ideas seems promising. 

Currently, a prototype implementation for the method sketched
in Sections~\ref{subsecMaxStates} and~\ref{subsecParamsForConsts}
is being developed
(for the class of programs which are at most tail recursive).
See
\begin{quote}
{\tt http://www.informatik.uni-halle.de/{\SN}brass/slddb/}.
\end{quote}

As a further generalization of the framework,
one could start subproofs for certain literals and
and possibly reuse their results multiple times,
in order to get magic sets, see~\cite{Bra00}.
It might also be possible to split goals in pieces
and link them together by references to previous states,
somewhat similar to~\cite{GZ92}.


\section*{Acknowledgements}


I would like to thank Heike Stephan for starting the research
on Earley deduction,
which inspired the abstraction presented here.
I would also like to thank Marcus Lehmann
for doing performance tests with the SLDmagic method.


\bibliographystyle{acmtrans}

\bibliography{../../ddb.bib}


\end{document}